\documentclass[aps,prd,fleqn,superscriptaddress]{revtex4}
\usepackage{graphicx,xcolor,natbib,braket}
\usepackage{amsmath,amssymb,amsfonts}

\newcommand{\bse}{\begin{subequations}}
\newcommand{\ese}{\end{subequations}}
\newcommand{\be}{\begin{equation}}
\newcommand{\ee}{\end{equation}}
\newcommand{\bea}{\begin{eqnarray}}
\newcommand{\eea}{\end{eqnarray}}
\newcommand{\ba}{\begin{array}}
\newcommand{\ea}{\end{array}}

\usepackage[colorlinks=true, linkcolor=blue, bookmarks=true]{hyperref}

\begin{document}

\title{HTEE vs. Pseudo-Entropy in Magnetic Fields}
%\author{ M. M. Daryaei Goki\footnote{$m_{-}$daryaeigoki@sbu.ac.ir}}
%\affiliation{Department of Physics, Shahid Beheshti University, 1983969411, Tehran, Iran}
\author{M. Ali-Akbari\footnote{$\rm{m}_{-}$aliakbari@sbu.ac.ir}}
\affiliation{Department of Physics, Shahid Beheshti University, 1983969411, Tehran, Iran}
%\author{A. Davody\footnote{davody@ipm.ir}}
%\affiliation{School of Particles and Accelerators, Institute for Research in Fundamental Sciences (IPM),
%P.O.Box 19395-5531, Tehran, Iran}
%\author{H. Ebrahim\footnote{hebrahim@ut.ac.ir}}
%\affiliation{Department of Physics, University of Tehran, North Karegar Ave., Tehran 14395-547, Iran}
%\affiliation{School of Physics, Institute for Research in Fundamental Sciences (IPM),
%P.O.Box 19395-5531, Tehran, Iran}
%\author{ L. Shahkarami\footnote{l.shahkarami@du.ac.ir}}
%\affiliation{School of Physics, Damghan University, Damghan, 41167-36716, Iran}
\begin{abstract}
We compare the holographic timelike entanglement entropy with the pseudo-entropy arising from a two-qubit quantum mechanical system. In this model, we consider transitions from an initial thermal state to a final thermal state at fixed temperature under the influence of an external magnetic field. Our findings highlight significant discrepancies between the two quantities, which display markedly different behavior.
\end{abstract}

\maketitle
\tableofcontents

\section{Introduction}
The AdS/CFT correspondence, and more generally gauge-gravity duality, establishes a strong-weak coupling relation between a gauge theory in $d$ dimensions and a gravity theory in $d+1$ dimensions, where the gauge theory lives on the boundary of the gravity theory \cite{Maldacena:1997re, Witten:1998qj, Aharony:1999ti, Casalderrey-Solana:2011dxg}. This implies that every physical quantity (or process) in the strongly coupled gauge theory admits a geometric dual description on the gravity side. To elucidate this duality, the gravitational counterparts of various gauge theory quantities have been extensively studied in the literature. A key feature of this duality is that it often provides simple geometric candidates for quantities that are otherwise difficult to compute directly. One of the most extensively studied quantities in this context over the past two decades is entanglement entropy.

Consider a Hilbert space for the entire physical system, denoted by $\mathcal{H}_{A+B}$, which can be factorized into two subsystems as $\mathcal{H}_A \otimes \mathcal{H}_B$. Assume that our physical system is described by a pure state $|\psi\rangle$ with the corresponding density matrix $\rho = |\psi\rangle\langle\psi|$. The reduced density matrix for subsystem $A$ is then defined by $\rho_A = \mathrm{Tr}_B(\rho)$, obtained by tracing over the degrees of freedom in subsystem $B$. The entanglement entropy associated with subsystem $A$ is given by $E = -\mathrm{Tr}_A (\rho_A \log \rho_A)$ \cite{Calabrese:2004eu, Casini:2009sr, Nishioka:2009un, Ling:2015dma, Klebanov:2007ws, Liu:2013iza}. This quantity measures the strength of quantum correlations between the two subsystems, or equivalently, the amount of information lost about subsystem $A$ when subsystem $B$ is traced out.
 
On the gravity side, a simple and elegant dual was first introduced by Ryu-Takayanagi (RT) \cite{Ryu:2006bv, Ryu:2006ef}. They proposed that the entanglement entropy associated with subsystem $A$ is obtained by extremizing the area of a codimension-two surface, called the RT surface, in the gravity background. This surface is required to be homologous to $A$, meaning its boundary coincides with the boundary of subregion $A$ on the asymptotic boundary of the geometry, converting a complex quantum field theory calculation into a tractable geometric problem. This proposal provides a simple way to compute entanglement entropy using gauge-gravity duality. For this reason, it has attracted significant attention over the past two decades and has passed numerous nontrivial tests, including agreement with known conformal field theory results.

A holographic proposal for computing pseudoentropy has recently been introduced and has been the subject of extensive discussion \cite{Mollabashi:2021xsd, Caputa:2024gve, Doi:2022iyj, Doi:2023zaf, Nakata:2020luh, Jena:2024tly, Takayanagi:2025ula, Afrasiar:2024ldn, Nunez:2025ppd, Nunez:2025puk, Zhao:2025zgm}. Consider a physical system described by two pure states, an initial state $\ket{\psi}$ and a final state $\ket{\phi}$, defined on the full Hilbert space $\mathcal{H}_{A+B}$. The transition matrix between these states is then given by
\begin{equation}\label{transition}
\mathcal{T}=\frac{\ket{\psi}\bra{\phi}}{\braket{\phi|\psi}},
\end{equation}
which is generally non-Hermitian when $\ket{\psi}\neq\ket{\phi}$. The reduced transition matrix for subsystem $A$ is obtained by tracing over $B$,
\begin{equation}
\mathcal{T}_A=\mathrm{Tr}_{B}(\mathcal{T}),
\end{equation}
and the pseudo-entropy is given by
\begin{equation}\label{pseudo}
S_{A}=-\mathrm{Tr}_{A}(\mathcal{T}_A\log\mathcal{T}_A).
\end{equation}
To compute pseudoentropy using gauge-gravity duality, one generalizes the RT prescription by considering timelike extremal surfaces instead of spacelike ones. This is achieved via analytic continuation and leads to what is called holographic timelike entanglement entropy (HTEE). For a two-dimensional conformal field theory, the HTEE obtained from this holographic proposal has been shown to agree with field theory results. This proposal has also been studied extensively in the literature, including its generalization to thermal states \cite{Caputa:2024gve, Gong:2025pnu} and its derivation via Wick rotation to obtain the same results \cite{Afrasiar:2024ldn}. Moreover, in \cite{Goki:2026hpl} it was argued that this proposal can reproduce similar behavior to that of a two-qubit system, although important differences were observed. The present work aims to further investigate these discrepancies.

In this paper, we investigate a two-qubit thermal system subjected to an external constant magnetic field. Following an analytic continuation, we compute the pseudoentropy for this system. We then employ the holographic proposal to evaluate the HTEE on the gravity side and compare the results obtained from both frameworks.

\section{Two-Qubit Toy Model}
In this section, we calculate the pseudo-entropy in a quantum mechanical setting to study its behavior in the presence of an external magnetic field. We consider a two-qubit system described by the following initial and final states.
The initial Hamiltonian is given by
\begin{equation}\label{Hi}
H_i = \sigma_1^x \otimes \sigma_2^x + \sigma_1^y \otimes \sigma_2^y + \sigma_1^z \otimes \sigma_2^z,
\end{equation}
where $\sigma^a$ are the Pauli matrices and $x, y, z$ denote spatial directions. This Hamiltonian is isotropic and exhibits an exchange symmetry $1 \leftrightarrow 2$. The initial density matrix is then
\begin{equation}\label{initial}
\rho_i = \frac{e^{-\beta H_i}}{\operatorname{Tr}\big(e^{-\beta H_i}\big)},
\end{equation}
where $\beta = T^{-1}$ and $T$ is the temperature of the thermal bath to which the two-qubit system is coupled.
The final Hamiltonian is taken as
\begin{equation}\label{Hf}
H_f = H_i + B_1\big(\sigma_1^z \otimes \mathbb{I}_2\big) + B_2\big(\mathbb{I}_2 \otimes \sigma_2^z\big),
\end{equation}
where $\mathbb{I}_2$ is the $2\times2$ identity matrix. Here $B_1$ and $B_2$ represent external magnetic fields applied separately to each qubit. It is clear that if $B_1 \neq B_2$, the exchange symmetry is broken and the two qubits are treated differently by the last two terms. However, when $B_1 = B_2$, the exchange symmetry is restored. The corresponding final density matrix is
\begin{equation}\label{final}
\rho_f = \frac{e^{-\beta H_f}}{\operatorname{Tr}\big(e^{-\beta H_f}\big)}.
\end{equation}

To compute the pseudo-entropy, we first construct the transition matrix from the initial state \eqref{initial} to the final state \eqref{final} as
\begin{equation}
\mathcal{T} = \frac{\rho_f \rho_i}{\operatorname{Tr}(\rho_f \rho_i)}.
\end{equation}
Note that this expression generalizes the transition matrix to mixed states and reduces to \eqref{transition} for two pure states. By tracing out subsystem $A$ or $B$ (in our case, qubit $1$ or $2$), we obtain the reduced transition matrices:
\begin{equation}
\mathcal{T}_A = \operatorname{Tr}_B(\mathcal{T}), \qquad
\mathcal{T}_B = \operatorname{Tr}_A(\mathcal{T}).
\end{equation}
Finally, the pseudo-entropy for subsystem $A$ is given by
\begin{equation}
S_A = -\operatorname{Tr}_A\big(\mathcal{T}_A \log \mathcal{T}_A\big),
\end{equation}
and similarly for subsystem $B$.

All calculations are detailed in Appendix~\ref{app}, where the eigenvalues of $\mathcal{T}_A$ and $\mathcal{T}_B$ are found to be
\begin{subequations}
\begin{align}
\lambda_{1A} &= \frac{1}{D}\big(a q_1 + D_1 A_1 + D_2 A_2\big), \\[4pt]
\lambda_{2A} &= \frac{1}{D}\big(a q_4 + D_2 A_2 + D_3 A_1\big),
\end{align}
\end{subequations}
and
\begin{subequations}
\begin{align}
\lambda_{1B} &= \frac{1}{D}\big(a q_1 + D_3 A_1 + D_2 A_2\big), \\[4pt]
\lambda_{2B} &= \frac{1}{D}\big(a q_4 + D_2 A_2 + D_1 A_1\big).
\end{align}
\end{subequations}
The pseudo-entropy for subsystem $A$ then reads
\begin{equation}\label{EE}
S_A = -\big(\lambda_{1A} \log \lambda_{1A} + \lambda_{2A} \log \lambda_{2A}\big),
\end{equation}
and analogously for $B$. It is evident that $S_A \neq S_B$. One can verify that $\lambda_{1A} = \lambda_{1B}$ and $\lambda_{2A} = \lambda_{2B}$ when $D_1 = D_3$, which corresponds to $B_1 = B_2$. That is, when the exchange symmetry is preserved, the pseudo-entropy is the same for both subsystems. Furthermore, in this case there is no imaginary part for the transition matrix associated with the pseudo-entropy and one can show that the eigenvalues of the transition matrix are both equal to $1/2$ for $B=0$.
To introduce an imaginary part into the transition matrix, we employ analytic continuation, as explained in Appendix~\ref{analytic}, and replace $B$ with $iB$. Unfortunately, one can show that although the eigenvalues become complex, the imaginary part of the pseudo-entropy vanishes for $B_1=B_2=iB$.
Since our goal is to compare quantum mechanical results with holographic calculations in the presence of an external magnetic field, we focus on the case $B_1 = iB$ and $B_2 = B$. This simplification makes the computations more tractable while still retaining the influence of the magnetic field, as seen, for example, in equations \eqref{ef1} and \eqref{ef4}.

\section{Holographic Framework}\label{holography}
The effects of an external constant magnetic field on various physical quantities 
have been extensively studied via gauge-gravity duality. 
To define a constant magnetic field on the $(3+1)$-dimensional gauge theory side, 
one considers on the gravity side a five-dimensional Einstein--Maxwell action 
with a negative cosmological constant. 
In order to describe a constant magnetic field, a suitable gauge field configuration 
is chosen such that a component of the field strength is constant; 
a standard choice is $A_y = B x$, yielding $F_{xy} = B$. 
Varying the action with respect to the metric and the gauge field yields the 
equations of motion which although straightforward in principle, are lengthy 
to solve explicitly. 
This procedure has been carried out extensively in the literature, leading to 
various asymptotically $AdS_5$ solutions in the presence of a constant magnetic field 
(see, e.g.,~\cite{DHoker:2009mmn}).
The key point we wish to emphasize is that the HTEE we study is sensitive to the entire radial direction, from the horizon deep in the bulk to the asymptotic boundary. 
Consequently, approximations based solely on near-horizon or near-boundary limits 
are insufficient and the full bulk solution must be employed.

We start with a general five-dimensional metric 
\begin{equation}
ds^2 = g_{tt}(r) \, dt^2 + g_{rr}(r) \, dr^2 + g_{xx}(r) \, dx^2 + g_{yy}(r) \, dy^2 + g_{zz}(r) \, dz^2,
\label{metricnew}
\end{equation}
where $r$ denotes the radial coordinate and all metric components are functions of $r$ alone. 
The horizon and the boundary are located at $r = r_h$ and $r = 0$, respectively. 
The horizon is characterized by the conditions $g_{tt}(r_h) = 0$ and $g^{rr}(r_h) = 0$, 
equivalently $g_{rr}^{-1}(r_h) = 0$. 
The coordinates $x$, $y$ and $z$ parametrize the three-dimensional spatial boundary.

The subregion \( A \) for which we aim to compute the pseudoentropy is defined as
\begin{equation}
A = \left\{ 
-\frac{L_t}{2} \leq t \leq \frac{L_t}{2}, \quad x = 0 
\right\},
\label{subregion}
\end{equation}
where \( L_t \) denotes the length of the time interval.
According to the RT prescription, the HTEE is obtained by extremizing the area of a codimension-two surface extending into the bulk. Parameterizing the surface by \( t(r) \) and using the metric \eqref{metricnew}, the area functional becomes
\begin{align}
S = \frac{V_{2}}{4G_N^{(5)}} \int_{r_i}^{r_f} dr \, 
\sqrt{ g_{yy} g_{zz} \bigl( g_{tt} t^{\prime 2} + g_{rr} \bigr) },
\label{area}
\end{align}
where \( V_{2} = \int dy \, dz \) is the (infinite) transverse volume and \( t' = dt/dr \). Treating the above expression as a Lagrangian that depends only on \( t'(r) \), the equation of motion for \( t(r) \) is given by
\begin{equation}
t'^{2} = \frac{c^{2} g_{rr}}{g_{tt} \bigl( g_{tt} g_{zz} g_{yy} - c^{2} \bigr)},
\label{55}
\end{equation}
where \( c \) is an integration constant. Substituting \eqref{55} into \eqref{area} yields
\begin{align}
S = \frac{V_{2}}{4G_N^{(5)}} \int_{r_i}^{r_f} dr \, g_{yy} g_{zz} \sqrt{\frac{g_{tt} g_{rr}}{g_{tt} g_{zz} g_{yy} - c^{2}}}.
\label{5}
\end{align}

In the standard black hole solution outside the horizon, we have \( g_{tt} = -|g_{tt}| < 0 \). For a real integration constant \( c \), the denominator in \eqref{55} never vanishes across the entire radial range from the horizon to the boundary. Consequently, there is no turning point \( r = r_* \) where \( t'(r) \) diverges meaning that the surfaces described by a real \( c \) do not turn back towards the boundary or horizon. To describe a surface that returns to the boundary or horizon, \( c \) must be chosen as a purely imaginary number, {\it{i.e.}} \( c = iC \). This choice yields a turning point at some \( r = r_* \) defined by the condition \( t'(r_*) \to \infty \). From \eqref{55}, this requires the denominator to vanish leading to \( C^2 = |g_{tt}(r_*)| \, g_{zz}(r_*) \, g_{yy}(r_*) \). By symmetry, the turning point is located at \( t = 0 \).

For a given background metric, one can use \eqref{55} with \( c = iC \) to determine \( t(r) \) numerically by integrating outward from the turning point with the condition \( t'(r_*) \to \infty \). In the region \( r \in (r_*, r_h) \), we have \( C^2 > |g_{tt}(r)| g_{zz}(r) g_{yy}(r) \), which causes the radicand in \eqref{5} to become negative. The area integral therefore evaluates to a purely imaginary number. We denote this as \( S = i S_I \). This solution connects the turning point to the horizon and gives the imaginary part of the pseudoentropy.

In addition, there exists a second class of solutions corresponding to the same magnitude \( C \) but with a real constant. These surfaces extend from a UV cutoff \( r = \epsilon \) near the boundary down to the horizon with the boundary condition \( t(\epsilon) = L_t/2 \). For this family of solutions, throughout the entire range \( r \in (\epsilon, r_h) \), the radicand in \eqref{5} is positive. The corresponding area integral is thus real and yields the real part of the pseudoentropy, {\it{i.e.}} \( S = S_R \). The full complex pseudoentropy is then given by \( S = S_R + i S_I \).

To compute $L_t$, we utilize equation \eqref{55}, obtaining
\begin{equation}
\frac{L_t}{2}=\int_{r_*}^{r_h} t'\big|_{c^2=-C^2} \, dr + \int_{r_h}^{\epsilon} t'\big|_{c^2=C^2} \, dr.
\end{equation}
Thus, both solutions discussed above contribute to $L_t$. The constant $C$ determines the value of $L_t$ at the boundary, and conversely, the boundary value determines $C$.

Since we aim to work with an analytical metric to carefully examine the results, we consider the following metric \cite{DHoker:2009mmn}
\begin{equation} \label{metric}
g_{tt}=-\frac{3}{r^2}\left(1-\frac{r^2}{r_h^2}\right), \quad 
g_{rr}=\frac{1}{3 r^2 \left(1-\frac{r^2}{r_h^2}\right)},\quad 
g_{xx}=g_{yy}=\frac{B}{\sqrt{3}},\quad 
g_{zz}=\frac{3}{r^2},
\end{equation}
and its corresponding temperature is given by
\begin{equation}
T=\frac{3 r_h}{2\pi}.
\end{equation}
Several key points should be noted about this metric:
\begin{itemize}
    \item This is an exact solution of the five-dimensional Einstein--Maxwell action 
with a negative cosmological constant and a non-zero constant magnetic field. It represents the product of a BTZ black hole and $T^2$ in the $x$ and $y$ directions.
    \item In the context of holography, this metric is considered to describe a quantum field theory in the large radial direction, or in other words, in the low-energy limit. As discussed in the original paper, a numerical solution has been found that interpolates between \eqref{metric} at low energy and $AdS_5$ at short distances.
    \item Note that $g_{xx}$ and $g_{yy}$ are constant and proportional to the external magnetic field which is oriented in the $z$-direction.
\item For this solution, the BTZ black hole radius is given by $l=\frac{L}{\sqrt{3}}$, for which we assume $L=1$ is the $AdS_5$ radius.
\end{itemize}

Applying the metric \eqref{metric}, the HTEE is obtained as 
\begin{align}\label{HTEE}
S & = \frac{V_{2}}{4G_N^{(5)}} \int_{r_i}^{r_f} dr \, \left(\frac{B}{\sqrt{3}}\right)^{1/2} g_{zz} \sqrt{\frac{g_{tt} g_{rr}}{g_{tt} g_{zz} \pm |g_{tt}^*| g_{zz}^* }},\\ \nonumber
&=\sqrt{B}\left[\frac{V_{2}}{4G_N^{(5)}} \int_{r_i}^{r_f} \frac{dr}{3^{1/6}r^2}\sqrt{\frac{-1}{-\left(1-\frac{r^2}{r_h^2}\right)\pm\frac{C^2}{9}r^4}}\right],
\end{align}
where $g_{tt}^*=g_{tt}(r_*)$, and the $\pm$ sign denotes the two families of solutions mentioned earlier. 
Two key observations regarding the HTEE are worth emphasizing:
\begin{itemize}
    \item As is evident from \eqref{HTEE}, both the real and imaginary parts of the HTEE scale as $\sqrt{B}$ with the external magnetic field in the low-energy regime of interest.
    \item In the limit $B \to 0$, the above expression becomes ill-defined. In this case, the HTEE must be recomputed directly from the metric \eqref{metric} with $B = 0$, which reproduces the known HTEE for the BTZ black hole reported in the literature.
\end{itemize}

\section{Results and Discussion}
\begin{figure}
\includegraphics[width=88 mm]{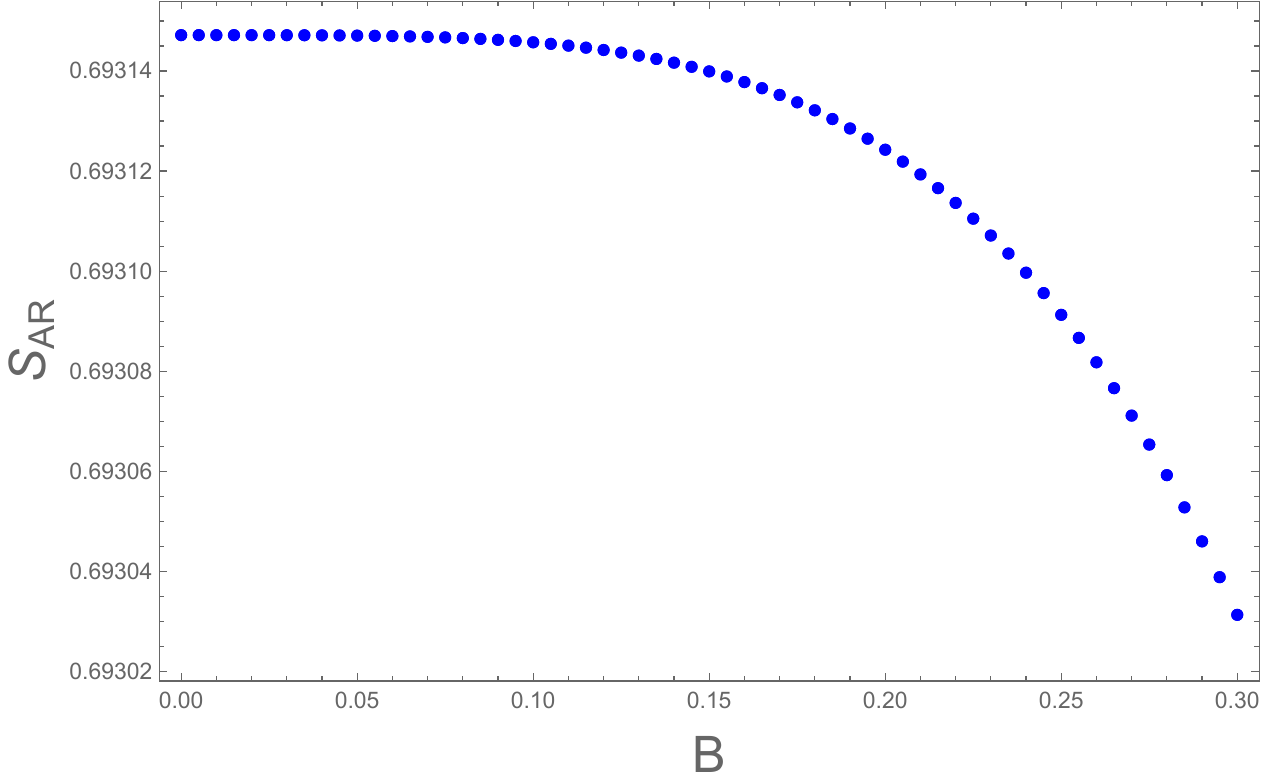}  
\includegraphics[width=88 mm]{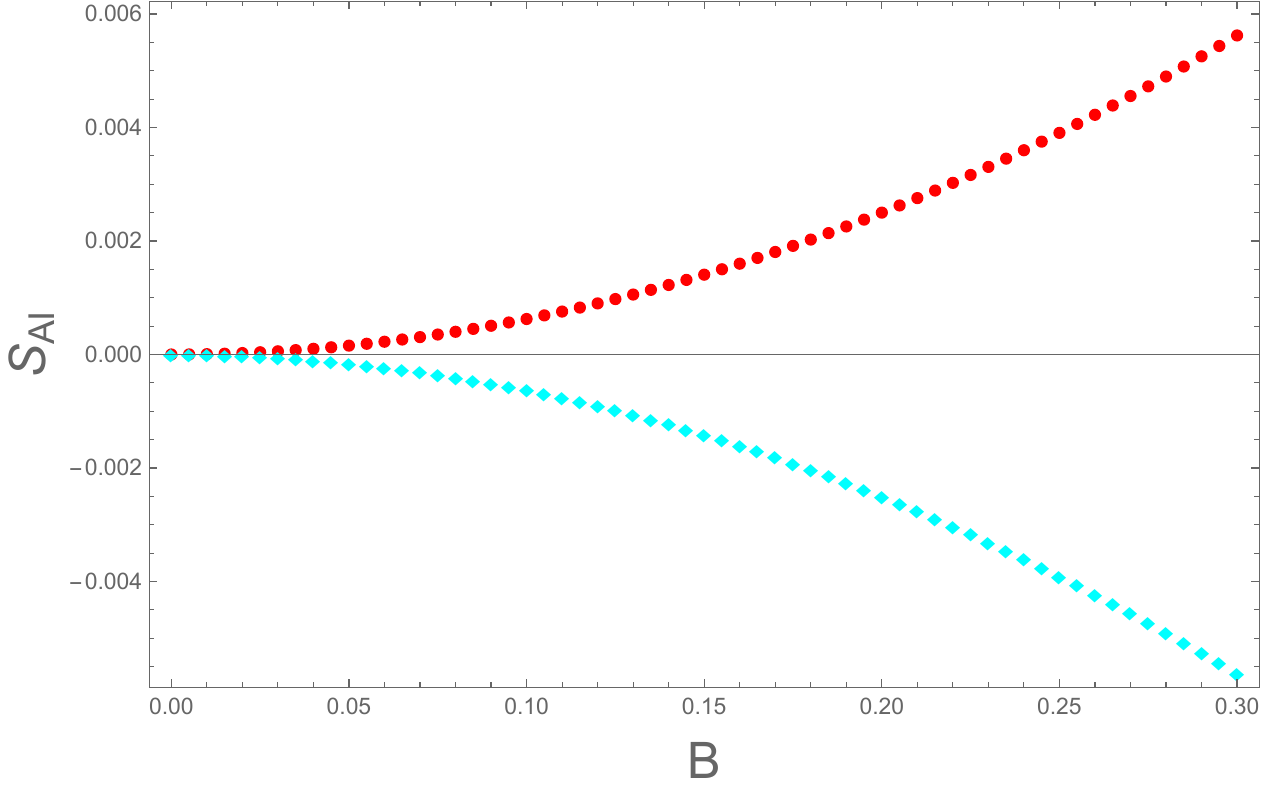}   
\includegraphics[width=88 mm]{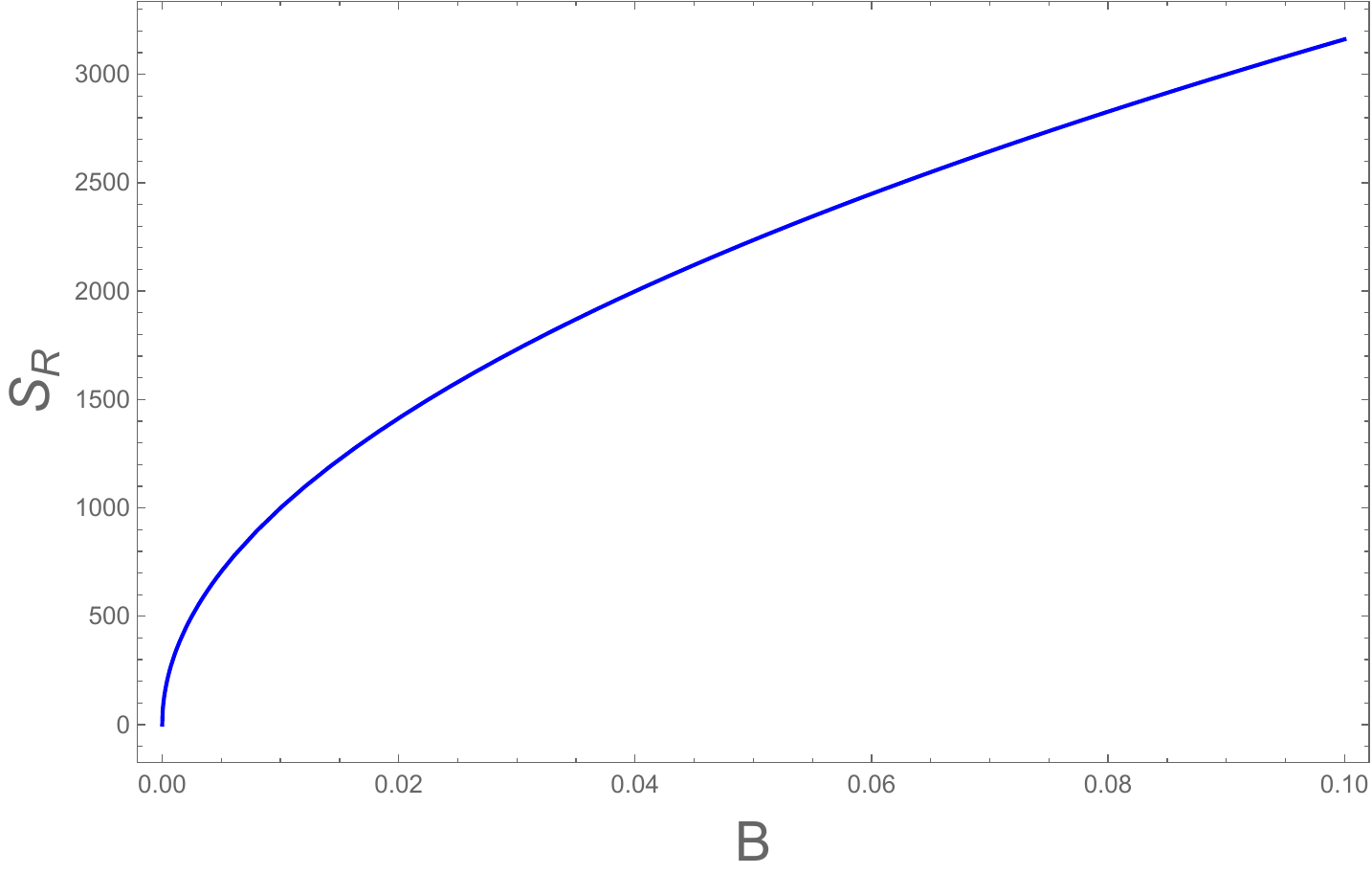}  
\includegraphics[width=88 mm]{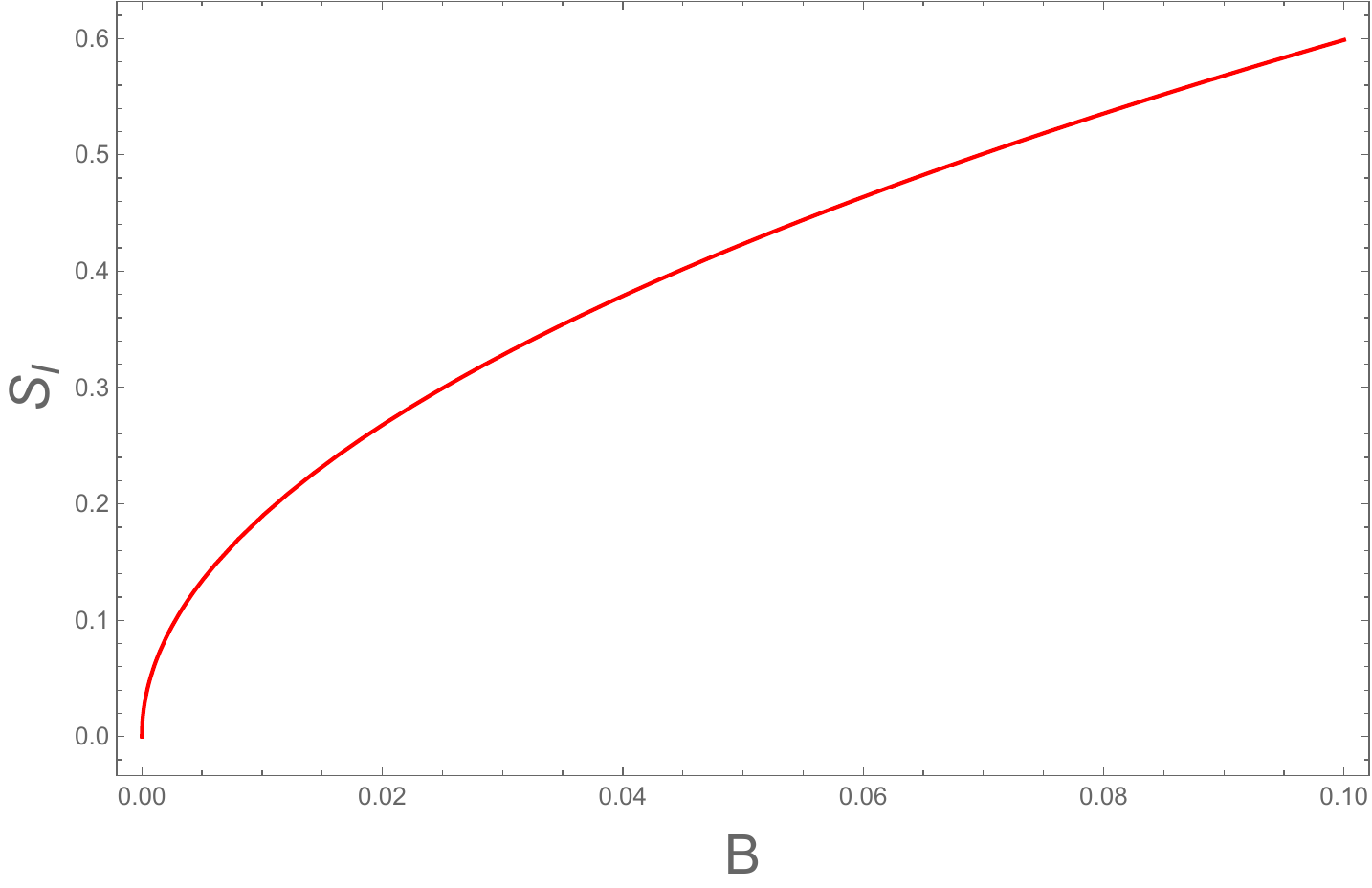}   
\caption{The real and imaginary parts of $S_{A}$ and $S\equiv \frac{4G_N^{(5)}}{V_2} S$ are plotted as functions of the external magnetic field $B$ for $T=0.143$, with $C=35$. The cyan curve corresponds to $-iB = B_1 = -i B_2$, meaning that replacing $i \rightarrow -i$ leaves $S_{AR}$ unchanged.}
\label{fig1}
\end{figure}

In this section, we compare our results for the transition matrix in quantum mechanics for a two-qubit system with the HTEE which has been proposed as a candidate to describe pseudoentropy. To facilitate this comparison, we expand the transition matrix for the case \(iB = B_1 = i B_2\) in the limit of small \(B\). This expansion yields

\begin{equation}\label{expansion}
S_A = \log 2 + \frac{B^2 \left[ i\left(1 - 2 e^{8 \beta} + e^{16 \beta} - 8 \beta^2\right) - 4 \sqrt{2} \beta + 4 \sqrt{2} e^{8 \beta} \beta \right]}{8 \left(1 + \sqrt{2} + \sqrt{2} e^{8 \beta}\right)^2} + \mathcal{O}(B^3).
\end{equation}
From this expression, we can identify the real and imaginary parts of the pseudoentropy as
\begin{align}
\label{Re:ps}
S_{AR} &= \log 2 + \frac{B^2 \left( \sqrt{2} \beta (e^{8 \beta} - 1) \right)}{2 \left(1 + \sqrt{2} + \sqrt{2} e^{8 \beta}\right)^2}, \\
\label{Im:ps}
S_{AI} &= \frac{B^2 \left( 1 - 2 e^{8 \beta} + e^{16 \beta} - 8 \beta^2 \right)}{8 \left(1 + \sqrt{2} + \sqrt{2} e^{8 \beta}\right)^2}.
\end{align}
The constant term \(\log 2\) appearing in the real part \(S_{AR}\) arises because, in the limit \(B \to 0\), the eigenvalues of the transition matrix become \(1/2\), resulting in a pseudoentropy of \(\log 2\). The remaining corrections to both the real and imaginary parts are proportional to \(B^2\).
This quadratic scaling with \(B\) stands in contrast to our findings from the holographic gauge theory calculation, where both the real and imaginary parts of the HTEE were found to scale as \(\sqrt{B}\). 
To further illustrate this point, in figure~\ref{fig1}, we present plots comparing the behavior of both the real and imaginary parts of the pseudoentropy from the two-qubit model with those of the HTEE from the holographic calculation. One can clearly observe the different behavior of the real and imaginary parts between the HTEE and the pseudoentropy.

Another point is that, according to \eqref{last}, we expect a constant term plus additional contributions arising from the magnetic field. However, it is important to note that in this paper we are working with quantum mechanics rather than quantum field theory and therefore it is reasonable that the finite-size term does not appear.

While the results from the simple quantum mechanical model cannot be expected to extend to the full holographic model, this discrepancy suggests that, at least within the specific framework we have studied here, the HTEE does not serve as an adequate candidate for pseudo-entropy. One can speculate as follows: The discrepancy between the scaling behaviors of HTEE ($\sqrt{B}$) and the quantum mechanical pseudo-entropy ($B^2$) likely stems from fundamental differences between the two frameworks. The quantum mechanical model is zero-dimensional with only four degrees of freedom, while the holographic calculation describes a $(3+1)$-dimensional CFT with infinitely many. Moreover, in holography the magnetic field backreacts on the geometry, modifying the metric itself, whereas in the quantum mechanical model it appears only as a Hamiltonian parameter. Whether a more sophisticated quantum mechanical model incorporating spatial structure or higher dimensions could reproduce the holographic scaling remains an open question for future investigation.
\appendix

\section{Detailed Calculations}\label{app}
Starting from \eqref{Hi}, using the basis ${\ket{00}, \ket{01}, \ket{10}, \ket{11}}$, the eigenvalues and eigenvectors of the initial Hamiltonian are easily obtained as
\begin{subequations}
\begin{align}
E_{i1}=+1;\ \ \ket{V_{i1}} &=\begin{pmatrix} 1 & 0 & 0 & 0 \end{pmatrix}^T, \\
E_{i2}=+1;\ \ \ket{V_{i2}} &=\begin{pmatrix} 0 & \frac{1}{\sqrt{2}} & \frac{1}{\sqrt{2}} & 0 \end{pmatrix}^T, \\
E_{i3}=+1;\ \ \ket{V_{i3}} &= \begin{pmatrix} 0 & 0 & 0 & 1 \end{pmatrix}^T,\\
E_{i4}=-3;\ \ \ket{V_{i4}} &= \begin{pmatrix} 0 & \frac{1}{\sqrt{2}} & -\frac{1}{\sqrt{2}} & 0 \end{pmatrix}^T.
\end{align}
\end{subequations}
The corresponding density matrix \eqref{initial} is therefore
%\begin{equation}
%a = e^{-\beta}Z_i^{-1}, \qquad b = e^{3\beta}Z_i^{-1}.
%\end{equation}

\begin{equation}
\rho_i = 
%\begin{pmatrix}
%a & 0 & 0 & 0 \\[4pt]
%0 & \dfrac{a+b}{2} & \dfrac{a-b}{2} & 0 \\[8pt]
%0 & \dfrac{a-b}{2} & \dfrac{a+b}{2} & 0 \\[8pt]
%0 & 0 & 0 & a
%\end{pmatrix}
\begin{pmatrix}
a & 0 & 0 & 0 \\[4pt]
0 & A_1 & A_2 & 0 \\[8pt]
0 & A_2 & A_1 & 0 \\[8pt]
0 & 0 & 0 & a
\end{pmatrix},
\end{equation}
where
\begin{align}
a &=Z_i^{-1}e^{-\beta}, \qquad
b = Z_i^{-1}e^{3\beta}, \\
A_1 &= \frac{1}{2}(a+b), \qquad
A_2 = \frac{1}{2}(a-b),
\end{align}
and $Z_i = \operatorname{Tr}\left(e^{-\beta H_i}\right) = 3e^{-\beta} + e^{3\beta}$. 

Similarly, the eiegnvalues and eigenvectors of the final Hamiltonian \eqref{Hf} are 
\begin{subequations}
\begin{align}
\label{ef1} E_{f1} &=+1 + B_1 + B_2, \qquad\qquad\ket{V_{f1}}=\begin{pmatrix} 1 & 0 & 0 & 0 \end{pmatrix}^T, \\
E_{f2} &= -1 + \sqrt{4 + |B_1 - B_2|^2}, \ \ket{V_{f2}}=N_2^{-1/2} \begin{pmatrix} 0 & 2 & \sqrt{4+|B_1-B_2|^2}-(B_1-B_2) & 0 \end{pmatrix}^T,\\
E_{f3} &= -1 - \sqrt{4 + |B_1 - B_2|^2}, \ \ket{V_{f3}}=N_3^{-1/2} \begin{pmatrix} 0 & 2 &-\left[\sqrt{4+|B_1-B_2|^2}+(B_1-B_2)\right] & 0 \end{pmatrix}^T,\\
\label{ef4} E_{f4} &=+1 - B_1 - B_2, \qquad\qquad\ket{V_{f4}} = \begin{pmatrix} 0 & 0 & 0 & 1 \end{pmatrix}^T,
\end{align}
\end{subequations}
with the nomalization canstants
\begin{subequations}
\begin{align}
N_2&=\sqrt{4+[\sqrt{4+|B_1-B_2|^2}-(B_1-B_2)]^2},\\
N_3&=\sqrt{4+[\sqrt{4+|B_1-B_2|^2}+(B_1-B_2)]^2}.
\end{align}
\end{subequations}
The density matrix \eqref{final} is 
\begin{equation}
\rho_f = 
%\begin{pmatrix}
%a & 0 & 0 & 0 \\[4pt]
%0 & \dfrac{a+b}{2} & \dfrac{a-b}{2} & 0 \\[8pt]
%0 & \dfrac{a-b}{2} & \dfrac{a+b}{2} & 0 \\[8pt]
%0 & 0 & 0 & a
%\end{pmatrix}
\begin{pmatrix}
q_1 & 0 & 0 & 0 \\[4pt]
0 & D_1 & D_2 & 0 \\[8pt]
0 & D_2 & D_3 & 0 \\[8pt]
0 & 0 & 0 & q_4
\end{pmatrix},
\end{equation}
where
\begin{align}
q_1&=Z_f^{-1} e^{-\beta E_{f1}},\qquad\ \ \ q_2=Z_f^{-1} N_2^{-2} e^{-\beta E_{f2}},\nonumber\\
q_3&=Z_f^{-1} N_3^{-2} e^{-\beta E_{f3}},\ \ \ q_4=Z_f^{-1} e^{-\beta E_{f4}},
\end{align}
with $Z_f\equiv\operatorname{Tr}\left(e^{-\beta H_f}\right) = e^{-\beta E_{f1}} + e^{-\beta E_{f2}} + e^{-\beta E_{f3}} + e^{-\beta E_{f4}}$ and 
\begin{align}
D_1&=4(q_2+q_3),\\
D_2&=2 \bigg[\left(\sqrt{4+|B_1-B_2|^2}-(B_1-B_2)\right) q_2-\left(\sqrt{4+|B_1-B_2|^2}+(B_1-B_2)\right) q_3\bigg],\\
D_3&=|\sqrt{4+|B_1-B_2|^2}-(B_1-B_2)|^2 q_2+|\sqrt{4+|B_1-B_2|^2}+(B_1-B_2)|^2 q_3.
\end{align}
The reduced transition matrices for sbusystems $A$ and $B$ are obtained as
\begin{equation}
{\cal{T}}_A=\operatorname{Tr}_B({\cal{T}})=\frac{1}{D}
\begin{pmatrix}
a q_1+D_1 A_1+D_2 A_2 & 0 \\[4pt]
0 & a q_4+ D_2 A_2+D_3 A_1  \\[8pt]
\end{pmatrix},
\end{equation}
with $D\equiv\operatorname{Tr}(\rho_f\rho_i)=a(q_1+q_4)+A_1(D_1+D_3)+2D_2 A_2$ and
\begin{equation}
{\cal{T}}_B=\operatorname{Tr}_A({\cal{T}})=\frac{1}{D}
\begin{pmatrix}
a q_1+D_3 A_1+D_2 A_2 & 0 \\[4pt]
0 & a q_4+ D_2 A_2+D_1 A_1  \\[8pt]
\end{pmatrix}.
\end{equation} 
Their corresponding eigenvalues are 
\begin{subequations}
\begin{align}
\lambda_{1A}&=\frac{1}{D}(a q_1+D_1 A_1+D_2 A_2),\\
\lambda_{2A}&=\frac{1}{D}(a q_4+ D_2 A_2+D_3 A_1),
\end{align}
\end{subequations}
and
\begin{subequations}
\begin{align}
\lambda_{1B}&=\frac{1}{D}(a q_1+D_3 A_1+D_2 A_2),\\
\lambda_{2B}&=\frac{1}{D}(a q_4+ D_2 A_2+D_1 A_1),
\end{align}
\end{subequations}
respectively.

\section{Analytic Continuation}\label{analytic}

We begin with the entanglement entropy for a two-dimensional conformal field theory
\be
S_{\rm{spacelike}}(L)=\frac{c}{3}\log{\frac{L}{\epsilon}},
\ee
where $L$ is the spacelike interval length and $\epsilon$ is the UV cutoff. Performing the analytic continuation $L\rightarrow i L_t$ yields:
\be\label{ACS}\begin{split}
S_{\rm{timelike}}(L_t)&=\frac{c}{3}\log{\frac{iL_t}{\epsilon}}\\ 
&=\frac{c}{3}\log{\frac{L_t}{\epsilon}}+\frac{i\pi c}{6}.
\end{split}\ee
Here, analytic continuation means we extend the function $S(L)$, originally defined for $L\in (0,\infty)$, to a function on $z\in \mathbb{C}$ minus the point zero, with an appropriate branch cut, and then evaluate it at $z = i L_t$. The imaginary part of \eqref{ACS} arises purely from the timelike nature of the interval.

In the presence of a magnetic field, two distinct analytic continuations are required:

\begin{itemize}
    \item \textbf{Geometric continuation}: $L \rightarrow iL_t$, which converts the spacelike interval length into a timelike duration. This provides the basic timelike structure.

    \item \textbf{Magnetic field continuation}: $B \rightarrow iB$ in our toy model, which distinguishes the initial and final states $\rho_i$ and $\rho_f$ through different magnetic field configurations. This ensures the reduced transition matrix is non-Hermitian with complex eigenvalues, thereby generating the state‑dependent imaginary part of the pseudoentropy.
\end{itemize}
The combined continuation $L \to iL_t$ and $B \to iB$ produces a complex‑valued pseudoentropy whose imaginary part encodes both the geometric timelike contribution and the quantum interference between states prepared with different magnetic fields.
More explicitly, the spacelike entanglement entropy in the presence of a magnetic field takes the form
\be 
S_{\rm{spacelike}}(L,B)=\frac{c}{3}\log{\frac{L}{\epsilon}} + f(B,L),
\ee
where $f(B,L)$ encodes the magnetic‑field dependence and possible finite‑size effects. Performing the double analytic continuation gives the pseudoentropy:
\be\label{pseudoACS}\begin{split}
S_{\mathrm{pseudo}}(L_t,iB)&=\frac{c}{3}\log{\frac{iL_t}{\epsilon}} + f(iB,iL_t) \\
&=\frac{c}{3}\log{\frac{L_t}{\epsilon}}+\frac{i\pi c}{6} + f(iB,iL_t).
\end{split}\ee
The continuation $B \to iB$ enters through $f(iL_t,iB)$, which contributes to both the real and imaginary parts. The total imaginary part is therefore
\be\label{last}
\mathrm{Im}\big[S_{\mathrm{pseudo}}(L_t,iB)\big] = \frac{\pi c}{6} + \mathrm{Im}\big[f(iL_t,iB)\big],
\ee
combining the universal geometric contribution $\pi c/6$ with terms that depend on the magnetic field.
It is not a priori clear whether the quantity obtained after a double analytic continuation in a simple model captures the same physics as the HTEE. Nevertheless, this identification seems reasonable and serves as a useful starting point for comparison.

\end{document}